\documentclass{article}

\PassOptionsToPackage{numbers, compress}{natbib}

    \usepackage[final]{neurips_data_2024}

\usepackage[utf8]{inputenc} 
\usepackage[T1]{fontenc}    
\usepackage{hyperref}       
\usepackage{url}            
\usepackage{booktabs}       
\usepackage{amsfonts}       
\usepackage{nicefrac}       
\usepackage{microtype}      
\usepackage{xcolor}         
\usepackage{bbding}
\usepackage{pifont}
\usepackage{amsthm}
\usepackage{amsmath}
\usepackage{multirow}
\usepackage{graphicx}
\usepackage{enumitem}
\usepackage{subfigure}
\usepackage{algorithm}
\usepackage{algorithmic}
\usepackage{threeparttable}

\theoremstyle{definition}
\newtheorem{definition}{Definition}

\title{A Large-scale Benchmark Dataset for Commuting Origin-destination Matrix Generation}

\author{%
	Can Rong$^1$
	\quad
	Jingtao Ding$^1$
	\quad
	Yan Liu $^{2}$
	\quad
	Yong Li$^1$\\
	$^1$Department of Electronic Engineering, Tsinghua University,
	\\
	$^2$Computer Science Department, University of Southern California. 
}

\begin{document}

\maketitle

\begin{abstract}
  The commuting origin-destination~(OD) matrix is a critical input for urban planning and transportation, providing crucial information about the population residing in one region and working in another within an interested area. Despite its importance, obtaining and updating the matrix is challenging due to high costs and privacy concerns. This has spurred research into generating commuting OD matrices for areas lacking historical data, utilizing readily available information via computational models. In this regard, existing research is primarily restricted to only a single or few large cities, preventing these models from being applied effectively in other areas with distinct characteristics, particularly in towns and rural areas where such data is urgently needed. To address this, we propose a large-scale dataset comprising commuting OD matrices for 3,233 diverse areas around the U.S. For each area, we provide the commuting OD matrix, combined with regional attributes including demographics and point-of-interests of each region in that area. We believe this comprehensive dataset will facilitate the development of more generalizable commuting OD matrix generation models, which can capture various patterns of distinct areas. Additionally, we use this dataset to benchmark a set of commuting OD generation models, including physical models, element-wise predictive models, and matrix-wise generative models. Surprisingly, we find a new paradigm, which considers the whole area combined with its commuting OD matrix as an attributed directed weighted graph and generates the weighted edges based on the node attributes, can achieve the optimal. This may inspire a new research direction from graph learning in this field.
\end{abstract}

\section{Introduction} \label{sec:Introduction}
Commuting refers to the daily round-trip movement of individuals from their homes to their workplaces. It is an important topic in fields like urban planning, transportation, environmental science, and economics~\cite{batty2007cities,gonzalez2008understanding,iqbal2014development,liu2020learning}. In form of the origin-destination~(OD) matrix, the commuting flow between every pair of regions within the interested area can be effectively recorded. We name it the commuting OD matrix, where each element represents the number of people reside in one region and work in another. The commuting OD matrix can be naturally modeled as a directed weighted graph, where nodes represent regions and edges represent the commuting OD flows between regions~\cite{saberi2017complex,saberi2018complex}. With the help of commuting OD matrices, urban planners can better understand the structured mobility patterns, optimize the transportation system, and make informed decisions on urban development~\cite{zeng2022causal,imai2021origin,zhong2014detecting}.

However, obtaining the commuting OD matrix is challenging. Traditionally, these matrices are collected through travel surveys or extracted from massive individual location records~(e.g., call detail records), which are expensive, time-consuming, and privacy-sensitive~\cite{rong2023interdisciplinary,wang2019origin}. To deal with this issue, researchers have proposed to generate commuting OD matrices using other readily available information~(e.g., sociodemographics and point-of-interests) via computational models~\cite{simini2021deep,rong2023complexity,rong2023goddag,pourebrahim2019trip,liu2020learning} to lower the barriers to data access. This task is referred to as commuting OD matrix generation. As the advancement of machine learning, leveraging large-scale datasets to train models has become a common practice in various fields, such as computer vision~\cite{dosovitskiy2020image} and natural language processing~\cite{radford2018improving}, no exception for the commuting OD matrix generation. So far, several models have been proposed to solve this problem, including classical physical models~\cite{zipf1946p,simini2012universal}, pairwised machine learning models~\cite{pourebrahim2019trip,liu2020learning,simini2021deep,rong2023goddag}, and deep generative models~\cite{rong2023complexity,rong2023origin}.

\begin{table}[t!]
    \caption{Comparison of the proposed dataset and other dataset utilized in existing works.}\label{tb:limitation}
    \begin{threeparttable}
    \resizebox{\columnwidth}{!}{
        \begin{tabular}{c|ccccccc}
        \hline
        Dataset                                            & \#Area   & Area Type         & Cover Area~(km$^2$) & Metropolitan & Town      & Rural      & Curated \& Public  \\ \hline
        Karimi et al.~\cite{karimi2020origin}              & 1        & Central District  & -                   & \ding{51}    & \ding{55} & \ding{55}  & \ding{55}   \\
        Pourebrahim et al.~\cite{pourebrahim2018enhancing,pourebrahim2019trip} & 1        & Whole City  & 789   & \ding{51}    & \ding{55} & \ding{55}  & \ding{55}   \\
        Liu et al.~\cite{liu2020learning}                  & 1        & Whole City        & 789                 & \ding{51}    & \ding{55} & \ding{55}  & \ding{51}   \\
        Yao et al.~\cite{yao2020spatial}                   & 1        & Central District  & 900                 & \ding{51}    & \ding{55} & \ding{55}  & \ding{55}   \\
        Lenormand et al.~\cite{lenormand2015influence}     & 2        & Whole City        & 15,755              & \ding{51}    & \ding{55} & \ding{55}  & \ding{55}   \\
        Rong et al.~\cite{rong2023goddag,rong2023origin,rong2023complexity} & 8 & Whole City  & 25,954          & \ding{51}    & \ding{55} & \ding{55}  & \ding{51}   \\
        Simini et al.~\cite{simini2021deep}                & 2,911    & National Gridding Coverage     & 686,983             & \ding{51}    & \ding{51} & \ding{51}  & \ding{55}   \\ \hline
        Ours                                               & 3,233    & Census Area Coverage       & 9,372,610           & \ding{51}    & \ding{51} & \ding{51}  & \ding{51}   \\ \hline  
        \end{tabular}
    }
    \end{threeparttable}
\end{table}

As shown in Table~\ref{tb:limitation}, existing models have a significant limitation, which is that they are primarily learned from a single or few large developed cities, such as New York City~\cite{liu2020learning}, Chicago~\cite{rong2023complexity}, and Los Angeles~\cite{rong2023goddag}. These cities have unique characteristics, such as high population density, advanced economics, and complex transportation systems. The models trained on a handful of this kind of cities may not be generalizable to other areas with distinct characteristics, particularly in underdeveloped areas where commuting OD matrices are more urgently needed.

To break this limitation and facilitate the development of more univerisal models to solve the commuting OD matrix generation task, we collect data from multiple sources and construct a large-scale benchmark dataset containing commuting OD matrices for 3,233 diverse areas around the United States. For supporting better design of models, each area in the dataset has not only the commuting OD matrix but also regional sociodemographics and numbers of point-of-interests within different categories for all regions in the area. Specifically, each area is profiled with its boundary and the boundaries of regions within it, which are represented as polygons with detailed geographic coordinates, i.e., latitude and longitude. The sociodemographics include the population of different genders and age groups, the number of househoulds, and income levels, etc. The point-of-interests are categorized into various types, such as restaurants, education, and shopping, etc. 

\begin{figure}[t]
    \centering
    \includegraphics[width=0.8\textwidth]{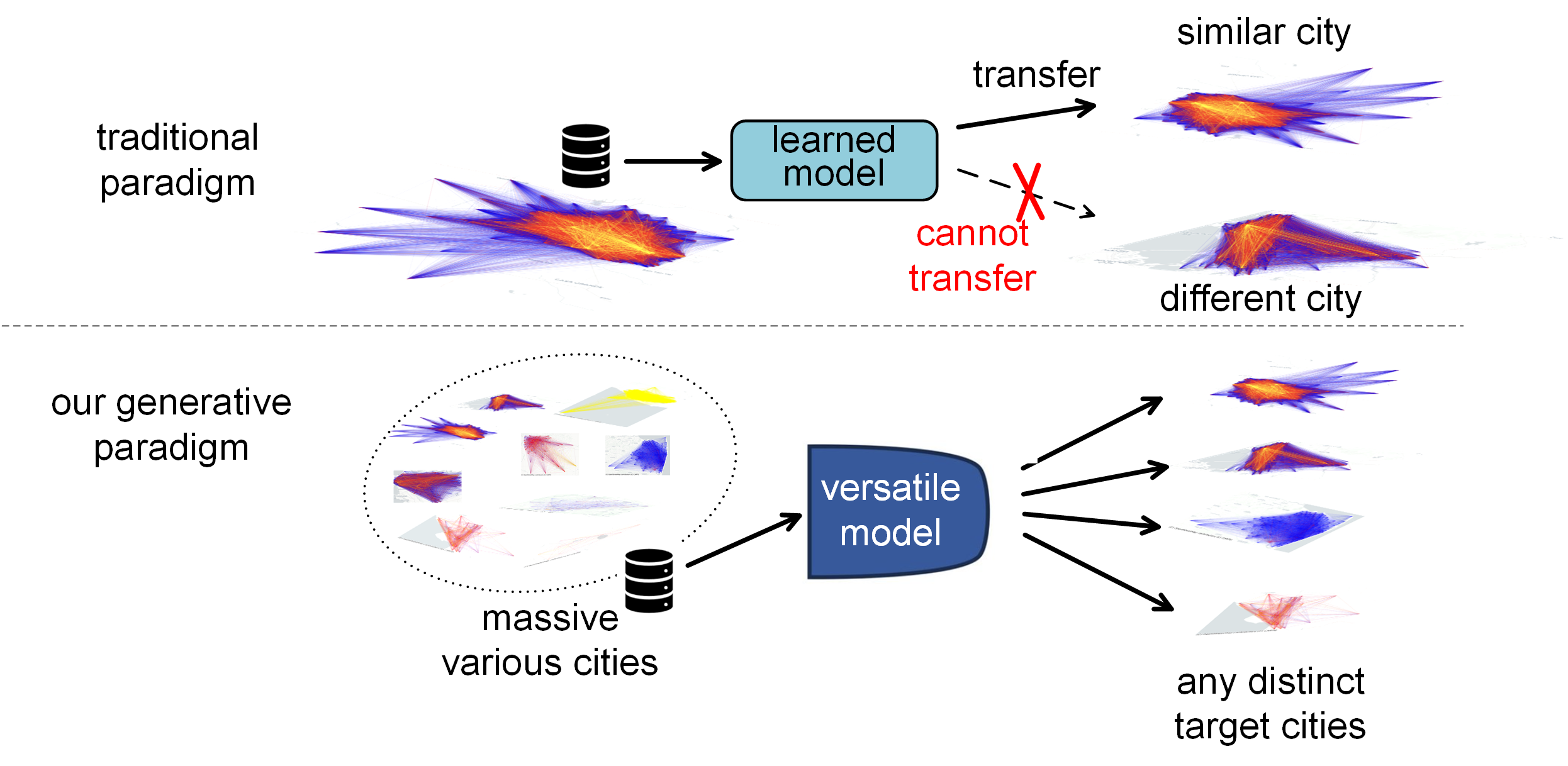}
    \caption{A comparison of traditional transfer paradigm and our novel generative paradigm for origin-destination matrix generation.}
    \vspace{-0.5cm}
    \label{fig:oneforall}
\end{figure}
Thanks to the extensive scale of the dataset, the geographic diversity of the areas enables the models to capture the commuting OD matrices of areas with different scales and structures. Compared to the datasets used in previous works, our dataset can establish a new paradigm for commuting OD matrix generation task. As shown in Figure~\ref{fig:oneforall}, the traditional transfer paradigm is to train a model on a single or few large cities and apply it to other target areas lacking commuting OD flow information. In contrast, our comprehensive dataset allows researchers to train models on a large number of diverse areas and generalize them to unseen areas. It is worth noting that the commuting OD matrix of an area can be viewed as a single graph data sample in the new paradigm and the commuting OD matrix generation task can be formulated as a conditional graph generation task, which generates the weighted edges of the graph based on the node attributes. This new paradigm will not only consider the local dependencies between the commuting OD flows and their corresponding origin and destination regions but also capture the global relational structure as a complex network~\cite{saberi2017complex,saberi2018complex} among all commuting flows within the area. 

We use the dataset to benchmark existing models and conduct preliminary investigation on performance of graph generation models in the new paradigm, highlighting their performance and limitations. All the models can be categorized into three types: physical models~\cite{zipf1946p,simini2012universal}, element-wise predictive models~\cite{pourebrahim2019trip,liu2020learning,rong2023goddag,simini2021deep}, and matrix-wise generative models~\cite{rong2023origin,rong2023complexity}. Surprisingly, we find that the matrix-wise generative models can achieve the optima. Enhanced by the training on massive diverse areas, the models can get better generalization performance. This may inspire a new research direction from the graph learning in the field of commuting OD matrix generation.

In summary, the contributions of this work are as follows:
\begin{itemize}[itemsep=2pt,topsep=2pt,parsep=0pt,leftmargin=*]
    \item We construct a large-scale benchmark dataset containing commuting OD matrices for 3,233 diverse areas around the United States for the commuting OD matrix generation task. Each area also includes regional sociodemographics and point-of-interests as attributes for the regions in the area.
    \item We propose a new paradigm for solving the commuting OD matrix generation task, which models the commuting OD matrix as a sample of an attributed directed weighted graph. The task can be formulated as a conditional graph generation task, which generates the weighted edges of the graph based on the attributed nodes.
    \item We benchmark the existing models and do preliminary investigation of the new paradigm. The results show that the matrix-wise generative models can achieve the optimal performance and the training on massive diverse areas can improve the models further, which give a new research direction from the graph perspective to solve the problem.
\end{itemize}
\section{Preliminaries} \label{sec:pre}
In this section, we introduce the definitions and problem formulation of the OD matrix generation task, followed by the existing works of this field.
\vspace{-0.2cm}
\subsection{Definitions and Problem Formulation}

\begin{definition}
    \textbf{Regions.} We divide the interested area into non-overlapping regions, represented as $\mathcal{R}=\{ r_i | i=1,2,...,N \}$, with $N$ being the total count of the regions. Each region fulfills unique functions, indicated by their characteristics $\mathbf{X}_r$, which include demographics and the distribution of points-of-interests in different categories.
\end{definition}
\begin{definition}
    \textbf{Spatial Characteristics.} The spaital characteristics of an area $ \mathcal{C_\mathcal{R}} $ are composed of the characteristics of each region $ \{\mathbf{X}_{r_i} | r_i \in \mathcal{R} \} $ and the interactions, such as distance, between all regions $ \{ d_{ij} | r_i \ \text{and} \ r_j  \in \mathcal{R} \} $ . 
\end{definition}
\begin{definition}
    \textbf{Commuting OD Flow.} The term commuting OD flow refers to the population $\mathcal{F}_{r_{org},r_{dst}}$, residing in $r_{org}$ and working at $r_{dst}$.
\end{definition}
\begin{definition}
    \textbf{Commuting OD Matrix.} Denoted by $\mathbf{F} \in \mathbb{R}^{N \times N}$, the commuting OD matrix includes commutings among all regions within the area. $F_{i,j}$ means the commuting from $r_i$ to $r_j$.
\end{definition}

\vspace{-0.2cm}
\textsc{Problem 1.} \textit{\textbf{Commuting OD Matrix Generation.} For any given areas, by specifying their spatial characteristics $ \mathcal{C}_\mathcal{R} $, generating their corresponding commuting OD matrices $\mathbf{F}$ that closely resemble those in the real world without any historical information.}

\vspace{-0.2cm}
\subsection{Existing Efforts on Commuting OD Matrix Generation Problem} \label{sec:Relatedwork}
Existing works can be categorized into three types. 
The \textbf{first} is classical physical models, such as the gravity model~\cite{zipf1946p} and the radiation model~\cite{simini2012universal}, which mimick the commuting OD flows as physical pheonomena and utilize simple mathematical equations to model the flows. 
The \textbf{second} is element-wise predictive models, such as tree-based models~\cite{robinson2018machine,pourebrahim2019trip}, support vector regression~(SVR)~\cite{rodriguez2021origin}, artificial neural networks~(ANNs)~\cite{sana2018using,lenormand2016systematic,simini2021deep}, which predict the OD flows between pairs of regions in data-driven schemes. Recently, as the research on OD flows from the perspective of complex networks deepens~\cite{saberi2017complex,saberi2018complex}, there is a growing trend of incorporating graph learning techniques~\cite{velivckovic2017graph,vignac2022digress,bojchevski2018netgan} to model the commuting OD matrix of a city. Some works~\cite{liu2020learning,yao2020spatial,rong2023goddag} utilized GNNs to model the urban spatial structure, predicting the missing part of the OD matrix. These semi-supervised graph learning methods cannot generalize to new cities without any commuting information. 
The \textbf{third} is matrix-wise generative models, such as the work by Rong et al.~\cite{rong2023origin,rong2023complexity} introducing adversarial and denoising diffusion generative methods to model the commuting OD matrix generation as graph generation problem. But they rely solely on data from a single city to train, limiting applications to generating OD matrices only for similar areas. All these works are based on the transfer paradigm shown in Figure~\ref{fig:oneforall}, limiting their generalization ability.
\section{Limitation of Dataset Used in Existing Works} \label{sec:limitation}

As shown in Table~\ref{tb:limitation}, existing datasets used in commuting OD matrix generation have several major limitations, which are detailed as follows:

\textbf{Limited Spatial Scale.} Existing datasets utilized in the literature have a limited spatial scale, usually focusing on a single or few large cities, leading two very limited spatial coverage. For example, Karimi et al.~\cite{karimi2020origin} and Yao et al.~\cite{yao2020spatial} only consider a central district in a city, and Pourebrahim et al.~\cite{pourebrahim2018enhancing,pourebrahim2019trip}, Liu et al.~\cite{liu2020learning}, Lenormand et al.~\cite{lenormand2015influence}, and Rong et al.~\cite{rong2023goddag,rong2023origin} only consider less than 8 large metropolitans, whose areas are less than 30,000 km$^2$. Although Simini et al.~\cite{simini2021deep} consider a national gridding coverage in the United Kingdom and Italy, the area is still limited to 686,983 km$^2$. Besides, they do not provide the curated dataset for public use, which cannot be used for further research. In contrast, our dataset covers 3,233 areas around the United States, a total area of 9,372,610 km$^2$, providing a much broader spatial scale. And our dataset is curated and publicly available, which can be found at https://github.com/tsinghua-fib-lab/CommutingODGen-Dataset.

\textbf{Homogeneous Area Type.} With the limited spatial scale, existing datasets usually focus on a single area type, such as metropolitan areas, central districts, or whole cities, which cannot include a massive areas with high diversity in terms of size and structure. Models trained on such datasets may not be generalized to other areas with different characteristics, limiting their applicability on only similar areas. Our dataset covers metropolitan areas, towns, and rural areas around the United States, providing a more comprehensive dataset for training and evaluating models. With the diversity of areas, models trained on our dataset can be more generalizable.
\section{Dataset} \label{sec:dataset}
\vspace{-0.2cm}
\subsection{Data Description} \label{sec:dataDescription} We have collected data from a total count of 3,233 areas around the United States, using counties as area boundaries and census tracts as region units in the corresponding area, each area including its regional spatial characteristics and OD matrix.

\textbf{Regional Spatial Characteristics.} Each region is characterized by demographics and urban functionalities, derived from American Community Survey~(ACS)~\cite{USCensusBureau2012} by the U.S. Census Bureau and the distribution of POIs from OpenStreetMap~\cite{OpenStreetMap}. Demographics include the population structure of a region based on age, gender, income, education, and other factors, encompassing a total of 97 dimensions. POIs are divided into 36 different categories. The distances between regions are calculated using the planar Euclidean distance between their centroids. 

\textbf{OD matrices.} We construct the OD matrices for all areas using data on commuting patterns from the 2018 Longitudinal Employer-Household Dynamics Origin-Destination Employment Statistics (LODES) dataset. These matrices represent aggregated commuting flows within areas. Each entry in an OD matrix denotes the count of individuals residing in one region and working in another, effectively mapping the commuting patterns of workers across different regions. The LODES dataset is widely used in existing works~\cite{liu2020learning,pourebrahim2019trip,pourebrahim2018enhancing}. In this dataset, the commuting information is aggregated by the cooperation and other kind of work units, which is more reliable and accurate than the individual commuting data. Therefore, in the data collection process, information has been ensured to be representative at a national scale, thus eliminating sampling errors. 

\vspace{-0.2cm}
\subsection{Data Statistics} \label{sec:dataStatistics} 

\begin{figure}[t]
    \centering
    \begin{minipage}[t]{0.65\textwidth}
        \centering
        \subfigure[\#Regions]{
            \label{fig:N_regions}
            \includegraphics[width=0.29\textwidth]{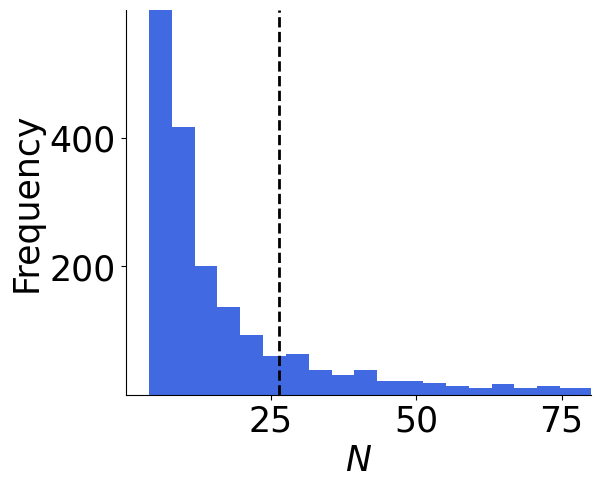}
        }
        \subfigure[Distances.]{
            \label{fig:avg_trip_dis}
            \includegraphics[width=0.29\textwidth]{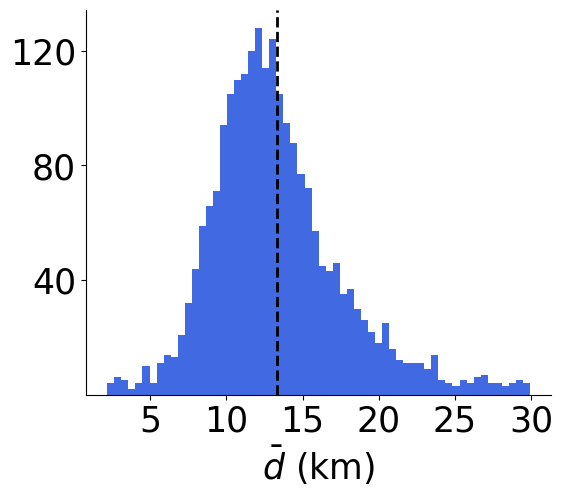}
        }
        \subfigure[Variance.]{
            \label{fig:var_degree}
            \includegraphics[width=0.29\textwidth]{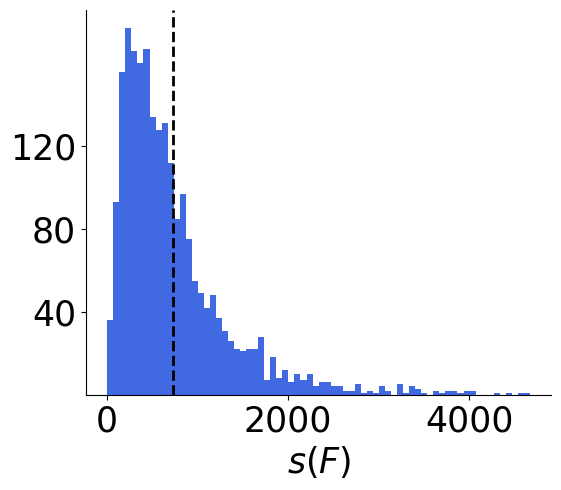}
        }
        \caption{Statistical analysis of the large-scale dataset of 3,233 areas in the United States, including the distribution of a) the number of regions in each area, b) the average trip distance in each area, c) the variance of the in/out flow of each region in each area.}
        \label{fig:data_statistics}

        \centering
        \subfigure[Maricopa]{
            \label{fig:OD_city_single}
            \includegraphics[width=0.29\textwidth, height=0.092\textheight]{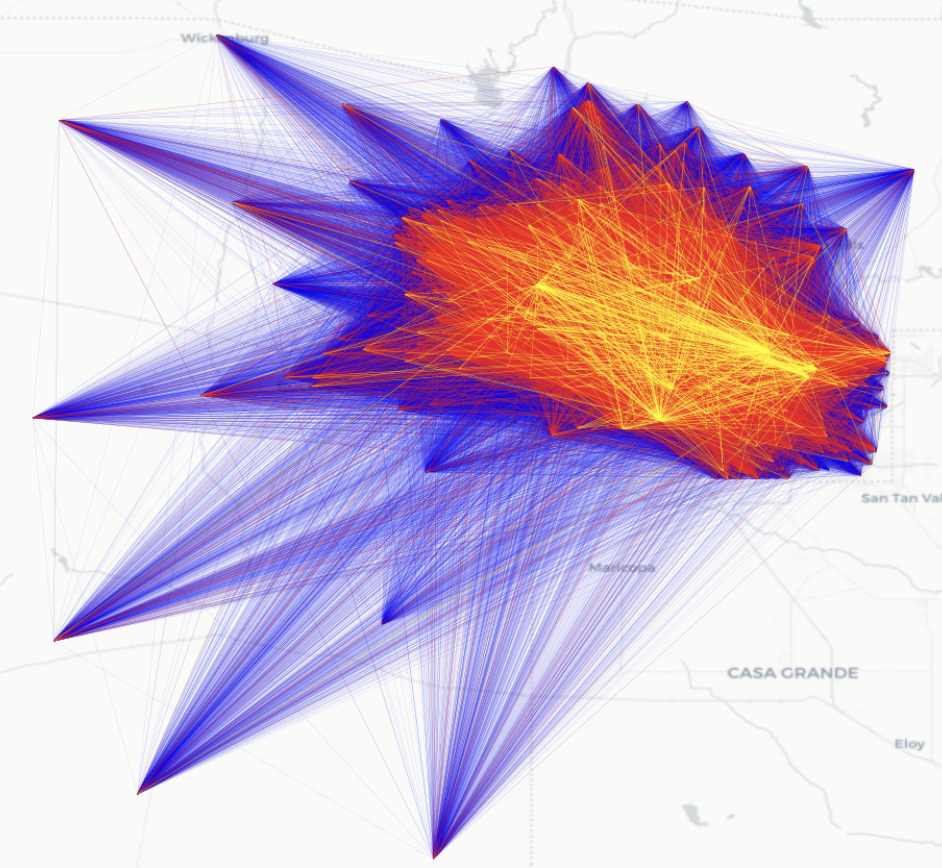}
        }
        \subfigure[Alameda]{
            \label{fig:OD_city_twopart}
            \includegraphics[width=0.29\textwidth, height=0.092\textheight]{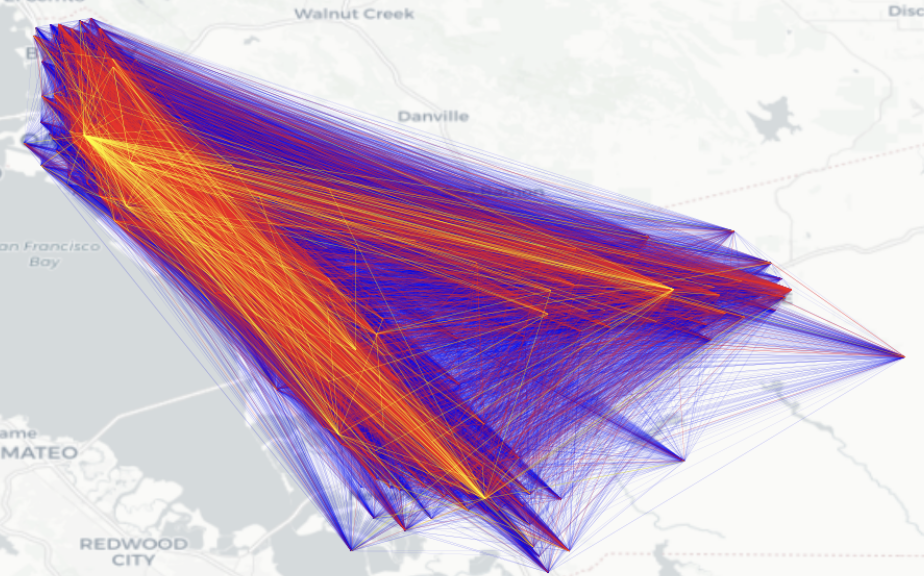}
        }
        \subfigure[Contra Costa]{
            \label{fig:OD_city_smooth}
            \includegraphics[width=0.29\textwidth, height=0.092\textheight]{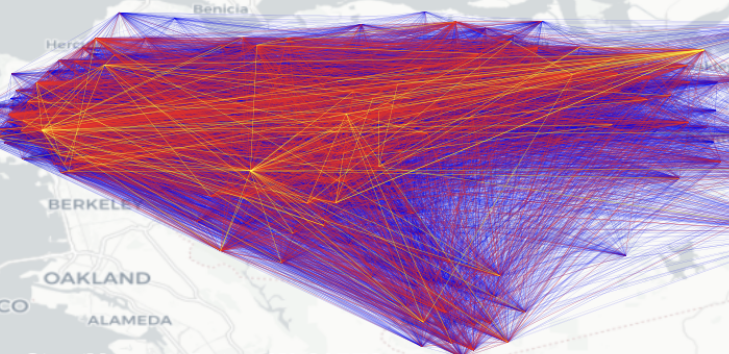}
        }
        \caption{Visualization of the OD matrices of three areas with different mobility structure, a) monocentric~(Maricopa in Arizona), b) polycentric~(Alameda in California), and c) smoothly distributed~(Contra Costa in California).}
        \label{fig:OD_viz_sample}

    \end{minipage}
    \hfill
    \begin{minipage}[t]{0.33\textwidth}
        \centering
        \subfigure[Edge weights.]{
            \label{fig:dist_edge_weight}
            \includegraphics[width=0.8\textwidth]{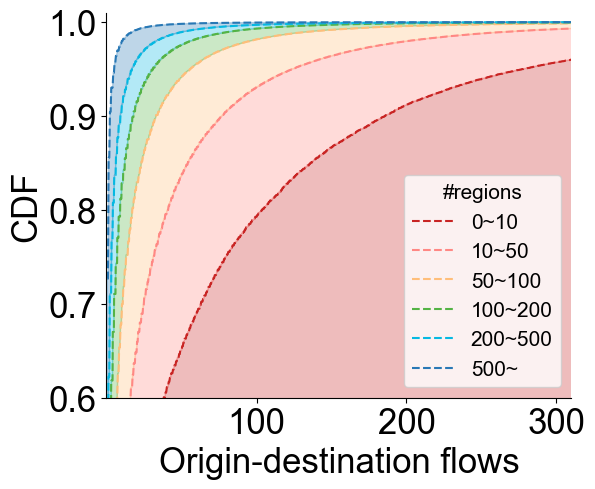}
        }
        \vspace{-0.2cm}
        \subfigure[Node degrees.]{
            \label{fig:dist_node_degree}
            \includegraphics[width=0.8\textwidth]{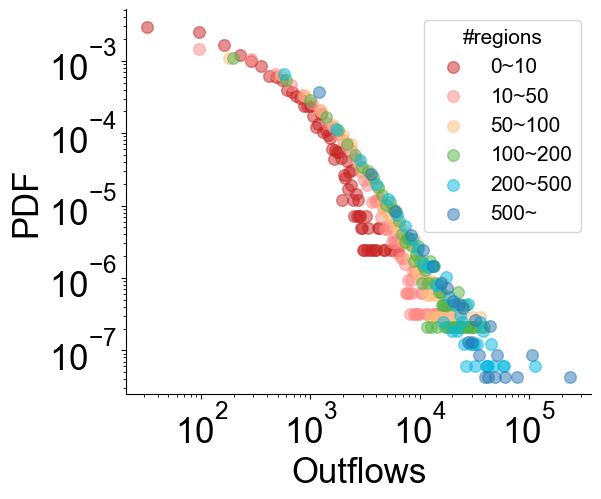}
        }
        \vspace{-0.2cm}
        \caption{Distributions of OD flows and outflows in areas of different scales. a) cumulative distribution function of edge weights, and b) probabilistic density function at log scale of node degrees.}
        \label{fig:skewness}
    \end{minipage}
\end{figure}

We conducted a visual analysis from three perspectives: number of regions, average trip distances, and variance of the regional mobility intensity. From Figure~\ref{fig:data_statistics}, it is evident that different areas exhibit significant variations in their structure and mobility patterns. Furthermore, Some cases shown in Figure~\ref{fig:OD_viz_sample} reveal the diverse patterns in mobility, including monocentric, polycentric, and evenly distributed mobility across different areas. Further, we analyze the distribution of OD flows and outflows in areas of different scales, as shown in Figure~\ref{fig:skewness}. We can observe that the heterogeneity exsits between different scales of areas. Yet, the commonalities also exist, i.e., the scaling behaviors are the same among areas.

\vspace{-0.2cm}
\subsection{Data Limitations} \label{sec:dataLimitations}

Despite the comprehensiveness of our dataset, there are several limitations that need to be addressed. First, the data is collected from a single year, which may not fully capture the temporal changes of commuting patterns. Second, the data is limited to the U.S., which may not be generalizable to other countries with different characteristics and cultures. Third, the data is aggregated at the county level, which may not fully capture the intra-county commuting patterns. We leave these limitations as future work to be addressed and maintain the dataset at the same time improve it.
\section{A New Paradigm and A Primary Exploration} \label{sec:methods}
\begin{figure*}[t]
    \centering
    \includegraphics[width=0.95\textwidth]{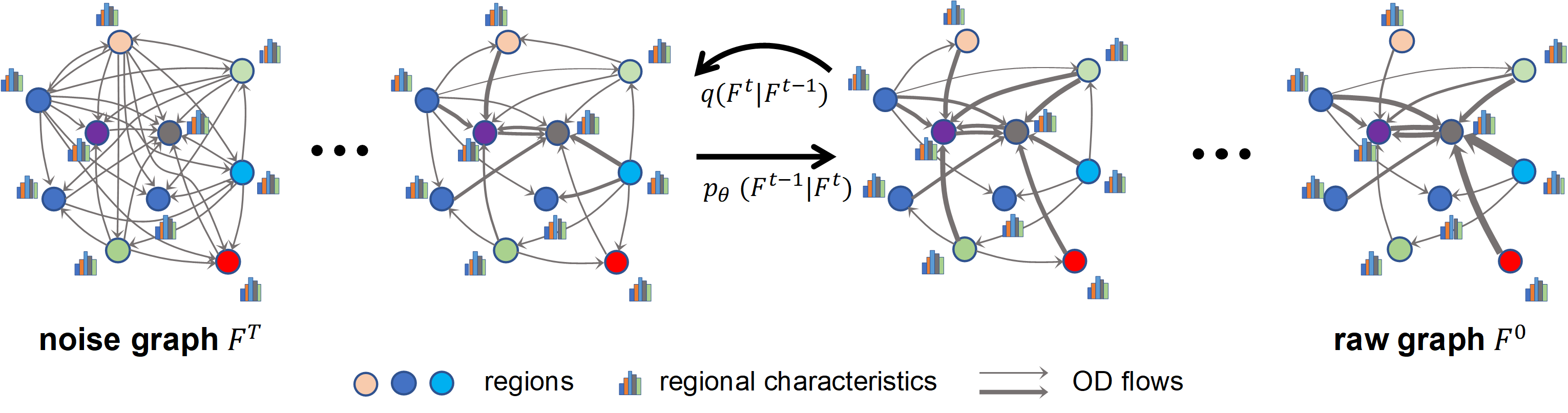}
    \vspace{-0.4cm}
    \caption{The framework of WeDAN for commuting OD matrix generation.}
    \vspace{-0.5cm}
    \label{fig:diffusion}
\end{figure*}

In this section, we give a detailed introduction to a new paradigm to solve the commuting OD matrix generation supported by our comprehensive dataset. In the new paradigm, we consider the whole area combined with its commuting OD matrix as an attributed directed weighted graph. Thus, the commuting OD matrix generation problem can be formulated as generating the weighted edges based on the attributed nodes. In this regard, we primarily adapt the graph generation model to the OD matrix generation task. To achieve better performance, we adopt the advanced diffusion-based graph generation model~\cite{vignac2022digress} to generate the weighted edges condition on the attributed nodes, which named WeDAN~(Weighted Edges Diffusion condition on Attributed Nodes). The framework of WeDAN is shown in Figure~\ref{fig:diffusion}. We will introduce the relevant the graph construction, diffusion process, denoising network, and the training and generation process in detail next.

\begin{figure}[t]
    \centering
    \includegraphics[width=0.65\textwidth]{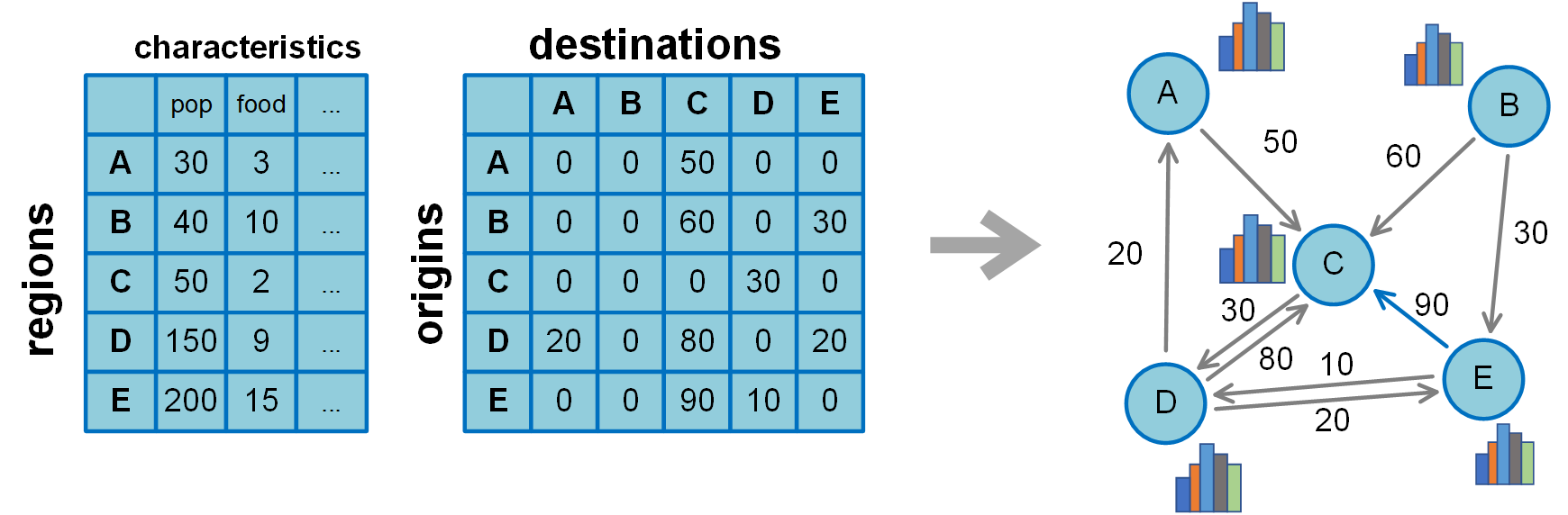}
    \caption{An example of construction of an attributed directed weighted graph formated by the spatial characteristics and OD matrix of the corresponding area consisting of 5 regions.}
    \vspace{-0.3cm}
    \label{fig:attrgraph}
\end{figure}

\textbf{Graph Construction.} As shown in Figure~\ref{fig:attrgraph}, we model an whole area as a graph $G=(\mathcal{V}, \mathcal{E})$. Specifically, each node $v \in \mathcal{V}$ on the graph represents a region $r$ within that area, and the directed edges $e_{ij} \in \mathcal{E}$ signify the commuting OD flows $\mathcal{F}_{r_{i},r_{j}}$ between regions. Herein, we let $N = |\mathcal{V}|$ be the number of nodes in the graph, representing the number of regions, where $|\dot|$ denotes the cardinality of a set. Each edge corresponds to its unique origin node and destination node. The weight of each edge $w_{e_{ij}}$ is the OD flow volume $F_{ij}$. The graph is attributed with the spatial characteristics of each region $r$, which are represented as the node features $\mathbf{X}_v$ of each node. The graph construction process is illustrated in Figure~\ref{fig:attrgraph}. Thus, the spatial characteristics of an area $\mathcal{C}_\mathcal{R}$ can be represented by a feature matrix $\mathbf{X}_\mathcal{R}$ composed of the attributes of all nodes $\{ v_r | r \in \mathcal{R} \}$ on the corresponding graph $G$, combined with the distances $ \{ d_{ij} | r_i \ \text{and} \ r_j  \in \mathcal{R} \} $ between all pairs of regions. Meanwhile, the commuting OD matrix $\mathbf{F}$ is equivalent to the set of all edges $ \{e | e \in \mathcal{E}\} $ and their weights $\{w_e | e \in \mathcal{E} \}$ on its graph $G$. 

By constructing a conditional generative model $\mathcal{P}_\theta(\mathcal{E},\{ w_e | e \in \mathcal{E} \} | \mathcal{V}, \mathbf{X}_\mathcal{R})$ that, given all nodes $\mathcal{V}$ and their attributes $\mathbf{X}_\mathcal{V}$ of a graph, generates all edges $\mathcal{E}$ and the corresponding weights $\{w_e | e \in \mathcal{E} \}$ on those edges, we can build an OD matrix generation model $\theta$. The conditional distribution $\mathcal{P}_\theta(\mathcal{E},\{ w_e | e \in \mathcal{E} \} | \mathcal{V}, \mathbf{X}_\mathcal{R})$ mirrors $ \mathcal{P}_\theta ( \mathbf{F} | \mathcal{C} ) $. 

\textbf{Weighted Edges Diffusion Condition on Attributed Nodes.} We will give a detailed introduction to the framework of the weighted edges diffusion process, which models the conditional distribution $\mathcal{P}_\theta(\mathcal{E},\{ w_e | e \in \mathcal{E} \} | \mathcal{V}, \mathbf{X}_\mathcal{R})$. As shown in Figure~\ref{fig:diffusion}, the diffusion framework is composed of two parts: the forward diffusion process $q$ and the reverse denoising process $p_\theta$. Both processes take place within the space of the edges $\mathcal{E}$ and the corresponding weights $ \{ w_e | e \in \mathcal{E} \} $ belong to the constructed attributed directed weighted graph. Since OD flows are numerical values, we employ continuous Gaussian noise in DDPM~\cite{ho2020denoising}.

Since OD matrices $\mathbf{F} \in \mathbb{R}^{N \times N} $ is similar to images, our forward diffusion process is also similar to DDPM~\cite{ho2020denoising}. The forward diffusion process is shown in Figure~\ref{fig:diffusion} from the right to the left. It is important to note that the noise perturbations applied to all edges are independent. So the forward diffusion process can be described at the individual OD flow level by the following computational formula.
\begin{equation}\label{eq:ODdiff}
    \begin{split}
    \begin{aligned}
    & q(F_{ij}^t|F_{ij}^{t-1})=\mathcal{N}(F_{ij}^t; \sqrt{1-\beta_t} F_{ij}^{t-1} , \beta_t \mathbf{I}),\\
    & q(F_{ij}^1, ..., F_{ij}^T|F_{ij}^0) = \prod_{t=1}^T {q(F_{ij}^t|q^{t-1})}.
    \end{aligned}
    \end{split}
\end{equation}
The reverse denoising process is the inverse of the forward diffusion process. In this context, the denoising process is facilitated by a denoising neural network $\theta$, which predicts the small noise $\epsilon$ to be removed based on the latent state of the noise space at step $t$, aiming to reach the noise state of step $t-1$, in an iteratively manner. Unlike the forward diffusion process, to ensure the modeling of the joint distribution of all elements in the OD matrix $\mathbf{F}$, the noise to be removed for each edge needs to be determined based on the entire state of the corresponding noisy data $\mathbf{F}^t$. Furthermore, to ensure the generation of OD matrices for new cities with given their spatial characteristics, we have designed the denoising process of OD matrices to be guided by the spatial characteristics of the corresponding cities, i.e., the nodes and their features. Therefore, the denoising step in reverse process can be represented as follows.
\begin{equation}\label{eq:ODdenoise}
    \begin{split}
    \begin{aligned}
    & p_{\theta} (\mathbf{F}^{t-1}|\mathbf{F}^t,\mathcal{C_\mathcal{R}}) = \mathcal{N} (\mathbf{F}^{t-1} ; \mu_{\theta}(\mathbf{F}^t,t,\mathcal{C_\mathcal{R}}) ,(1-\bar{\alpha}^t)\mathbf{I}),\\
    \end{aligned}
    \end{split}
\end{equation}
where
\begin{equation}\label{eq:miu_previous}
    \begin{split}
    \begin{aligned}
    & \mu_{\theta}(\mathbf{F}^t,t,\mathcal{C_\mathcal{R}}) = \frac{1}{ \sqrt{\alpha_t} } (\mathbf{F}^t - \frac{\beta_t}{\sqrt{1-\bar{\alpha}_t}} \mathbf{\epsilon}_\theta(\mathbf{F}^t,t,\mathcal{C_\mathcal{R}}) ) ,\\
    \end{aligned}
    \end{split}
\end{equation}
$ \alpha_t = 1-\beta_t$, and $\bar{\alpha}_t = \prod_{i=1}^t \alpha_i$. Here, $\mathbf{\epsilon}_\theta(\mathbf{F}^t,t,\mathcal{C_\mathcal{R}})$ is the noise predicted by $\theta$ based on the noisy state $\mathbf{F}^t$, diffusion step $t$ and the spatial characteristics $\mathcal{C_\mathcal{R}}$ of the corresponding city.

The denoising network $\theta$ is trained to predict the noise $\mathbf{\epsilon}_\theta(\mathbf{F}^t,t,\mathcal{C_\mathcal{R}})$ by minimizing the predictive errors. The parameterization and other detailed information of WeDAN is introduced in Appendix~\ref{apdx:paradigm}, such as architecture of the denoising network, algorithms of training and generation processes.
\section{Benchmark} \label{sec:exp}
\vspace{-0.2cm}
\subsection{Experimental Setup}
\textbf{Benchmark Models.} We utilize our propose dataset to benchmark ten existing models and our primiary exploration of the new paradigm, WeDAN. The existing models are in three categories: physical models, element-wise predictive models, and matrix-wise generative models. The physical models include classical gravity model~\cite{zipf1946p} with power-law decay~(GM-P) or exponential decay~(GM-E). The remaining models are data-driven approaches, including support vector regression~(SVR)~\cite{rodriguez2021origin}, random forest~(RF)~\cite{pourebrahim2018enhancing,pourebrahim2019trip}, gradient boosting regression tree~(GBRT)~\cite{robinson2018machine}, deep gravity model~(DGM)~\cite{simini2021deep}, geo-contextual multitask embedding learning~(GMEL)~\cite{liu2020learning} from the element-wise perspective and based on predictive schema. And the matrix-wise generative models include NetGAN~\cite{bojchevski2018netgan} and DiffODGen~\cite{rong2023complexity}. The details of the models are introduced in Appendix~\ref{apdx:baselines}.

\textbf{Evaluation Metrics.} We evaluate the performance from two perspectives: the error between the generated OD matrices and the corresponding real ones, and the distribution deviation in graph properties between the generation and the real data. The error metrics include Root Mean Square Error~({RMSE}), Normalized Root Mean Square Error~({NRMSE}) and Common Part of Commuting~({CPC}), while the distribution difference metrics include Jensen-Shannon Divergence~({JSD}) for inflow, outflow, and OD flow. These metrics are calculated for each area and then averaged across all. The calculation formulas are shown in Appendix~\ref{apdx:metrics}.

\begin{table}[t]
    \centering
    \caption{Performance comparison on all existing models.}
    \label{tab:performance}
    \resizebox{9.3cm}{!}{
        \begin{tabular}{@{\hspace{0.5pt}}c|ccc|ccc@{\hspace{0.5pt}}}
        \toprule
                & \multicolumn{3}{c|}{ Flow Value } & \multicolumn{3}{c}{ Property Distribution (JSD) }      \\ \midrule           
            \multicolumn{1}{c|}{Model}          & \multicolumn{1}{c}{CPC$\uparrow$} & \multicolumn{1}{c}{RMSE$\downarrow$} & \multicolumn{1}{c|}{NRMSE$\downarrow$} & \multicolumn{1}{c}{inflow$\downarrow$} & \multicolumn{1}{c}{outflow$\downarrow$} & \multicolumn{1}{c}{ODflow$\downarrow$}  \\ \midrule
        GM-P    & 0.321     & 174.0    & 2.222    & 0.668     & 0.656     & 0.409     \\
        GM-E    & 0.329     & 162.9    & 2.080    & 0.652     & 0.637     & 0.422     \\ \midrule
        SVR     & 0.420     & 95.4     & 1.218    & 0.417     & 0.555     & 0.410     \\ 
        RF      & 0.458     & 100.4    & 1.282    & 0.424     & 0.503     & 0.219     \\ 
        GBRT    & 0.461     & 91.0     & 1.620    & 0.424     & 0.491     & 0.233     \\ 
        DGM     & 0.431     & 92.9     & 1.186     & 0.469     & 0.561     & 0.230     \\
        GMEL    & 0.440     & 94.3     & 1.204     & 0.445     & 0.355     & 0.207     \\ \midrule
        NetGAN  & 0.487     & 89.1     & 1.138     & 0.429   & 0.354     & 0.191     \\
        DiffODGen & \underline{0.532}   & \underline{74.6}     & \underline{0.953}     & \underline{0.324}   & \underline{0.270}     & \underline{0.149}     \\ \midrule
        \multirow{2}{*}{WeDAN}   & \multicolumn{1}{c}{\multirow{2}{*}{\begin{tabular}[c]{@{}c@{}}\textbf{0.593}\\ \footnotesize{~(+11.5\%)}\end{tabular}}}    & \multicolumn{1}{c}{\multirow{2}{*}{\begin{tabular}[c]{@{}c@{}}\textbf{68.6}\\ \footnotesize{~(+8.04\%)}\end{tabular}}}     & \multicolumn{1}{c}{\multirow{2}{*}{\begin{tabular}[c]{@{}c@{}}\textbf{0.876}\\ \footnotesize{~(+8.04\%)}\end{tabular}}}     & \multicolumn{1}{c}{\multirow{2}{*}{\begin{tabular}[c]{@{}c@{}}\textbf{0.291}\\ \footnotesize{~(+10.2\%)}\end{tabular}}}   & \multicolumn{1}{c}{\multirow{2}{*}{\begin{tabular}[c]{@{}c@{}}\textbf{0.269}\\ \footnotesize{~(+0.96\%)}\end{tabular}}}     & \multicolumn{1}{c}{\multirow{2}{*}{\begin{tabular}[c]{@{}c@{}}\textbf{0.147}\\ \footnotesize{~(+1.34\%)}\end{tabular}}}     \\
                &           &          &           &           &           &          \\ \bottomrule
        \end{tabular}
        }
\end{table}

\vspace{-0.2cm}
\subsection{Performance Comparison}
The results are shown in Table~\ref{tab:performance}. All models utilize the ratio of 8:1:1 for dividing the data into training, validation, and test sets. We conducted experiments five times and averaged the results. The parameter settings are introduced in Appendix~\ref{apdx:parameters}.

\textbf{The exploration of the new paradigm, WeDAN, achieves the best performance.} The OD matrix generated by WeDAN demonstrates superior realism, from both flow value and property distribution deviation perspectives. Notably, in comparison to the top-performing baseline, WeDAN reduces RMSE/NRMSE by more than 8.0\% and improves the CPC over 11.5\%. Furthermore, the property distribution of the generated OD matrices closely matches the real ones, as evidenced by the lowest JSD from all the perspectives.

\textbf{The performance of data-driven approaches significantly outperforms the physical model.} The Gravity Model, using only four parameters, attempts to fit the complex human mobility, leading to inevitably underfitting. On the contrary, data-driven approaches, employing models with a multitude of parameters, go beyond by incorporating rich information such as demographics and POIs. Therefore, they have shown significantly better performance. 

\textbf{Modeling the joint distribution of all elements in OD matrices from the graph perspective hold advantages.} Modeling the dependency between the area's spatial space and the OD matrix globally, as opposed to merely modeling human flows between two regions~(i.e., origin and destination), results in a more effective capture of the properties of the mobility networks, i.e., OD matrices.

\textbf{Utilizing training data from various massive areas can enhance the performance.} Existing models based on graph generation have been designed only for large graphs, such as NetGAN and DiffODGen. In contrast, WeDAN is more versatile, capable of adapting to areas/graphs, of various sizes, from small to large. Consequently, it achieves more outstanding results.

\vspace{-0.2cm}
\subsection{Performance Analysis on Heterogeneous Areas}
To further explore the heterogeneity handled by the new paradigm and the applicability in different urban scenarios, we conducted comparative experiments on the model's performance across areas with various sizes and structures. Typically, developed areas are often larger and imply a stronger attraction to populations. Conversely, underdeveloped areas are usually small in size. Areas of different sizes also exhibit distinct mobility patterns, especially in terms of the skewness of OD flow distribution. Larger areas typically indicate stronger heterogeneity in mobility patterns from both node and edge perspectives, with a more pronounced long-tail effect in flow distribution.

\begin{figure}[t]
    \centering
    \subfigure[Areas of different sizes on CPC.]{
        \includegraphics[width=0.19\textwidth]{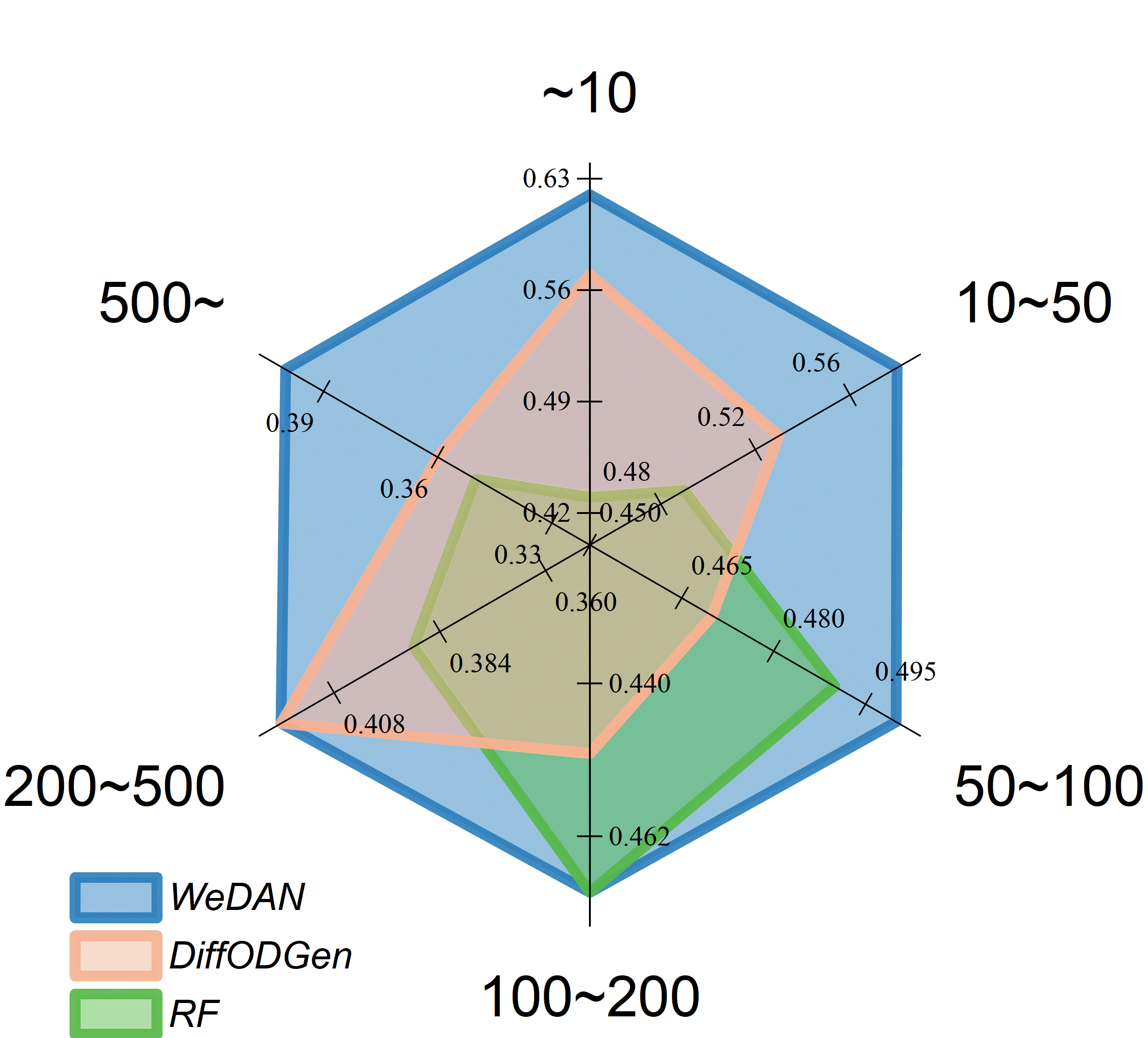}
    }
    \subfigure[Areas of different sizes on RMSE.]{
        \includegraphics[width=0.19\textwidth]{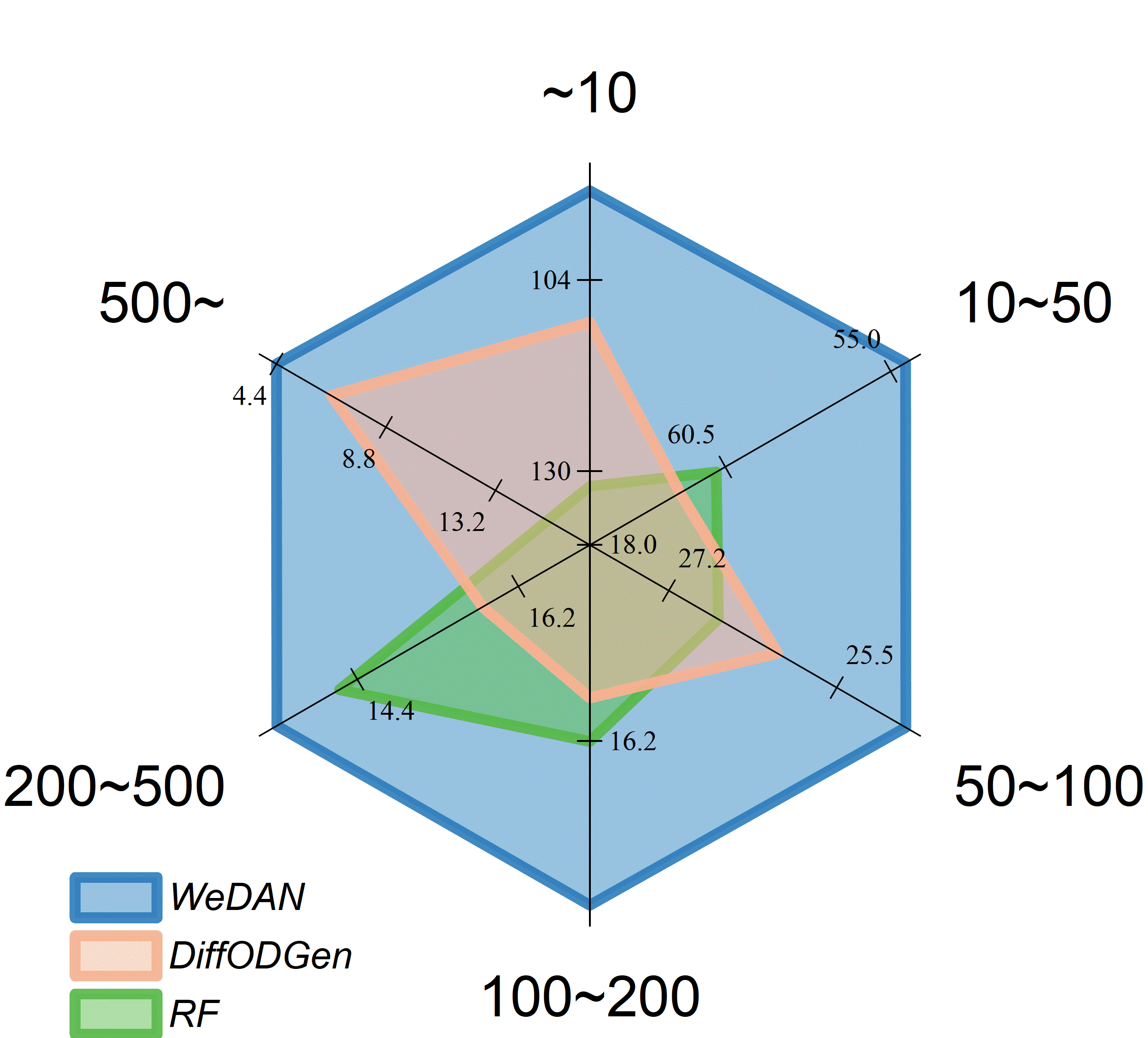}
    }
    \subfigure[Areas of different structures on CPC.]{
        \includegraphics[width=0.25\textwidth]{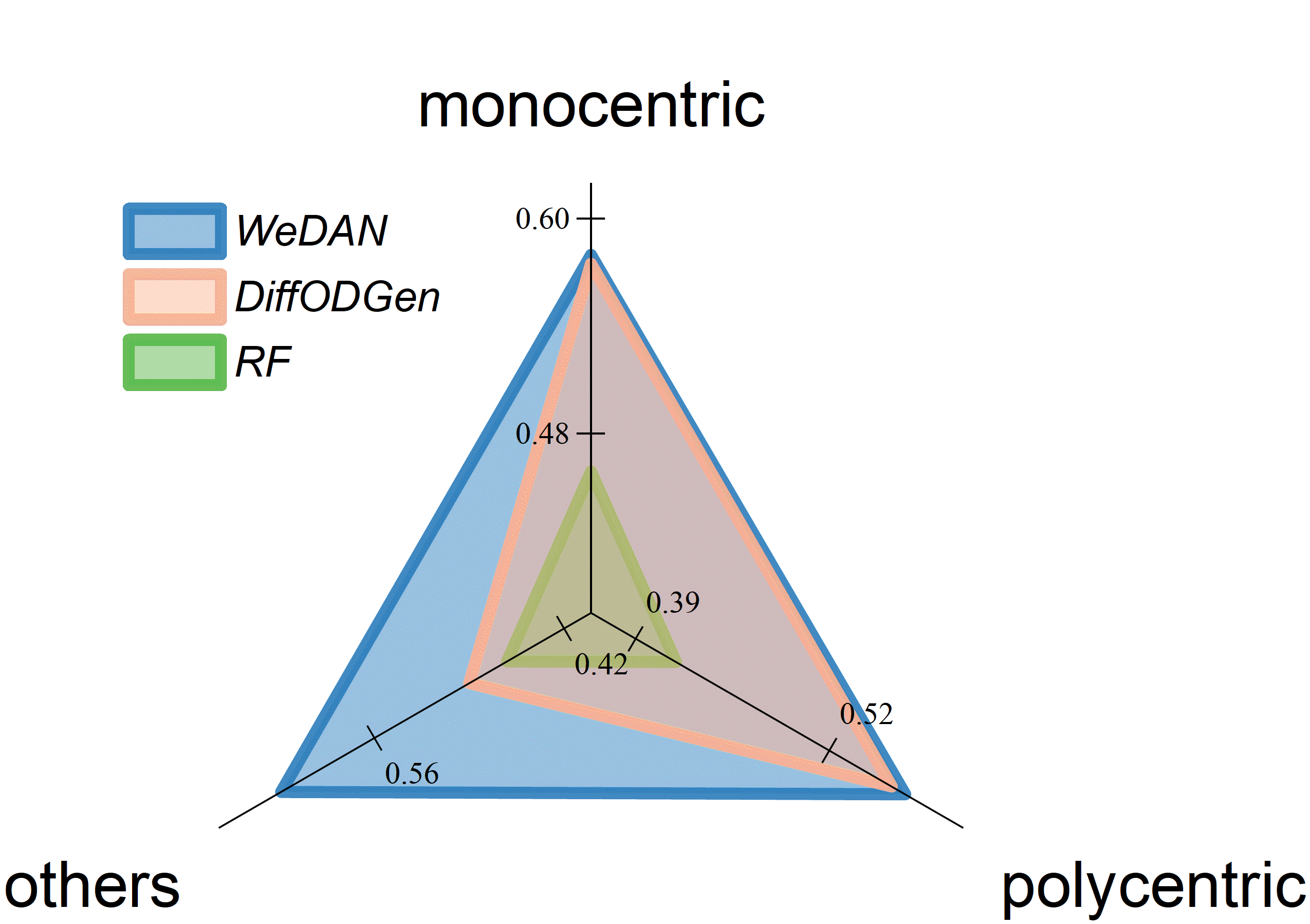}
    }
    \subfigure[Areas of different structures on RMSE.]{
        \includegraphics[width=0.25\textwidth]{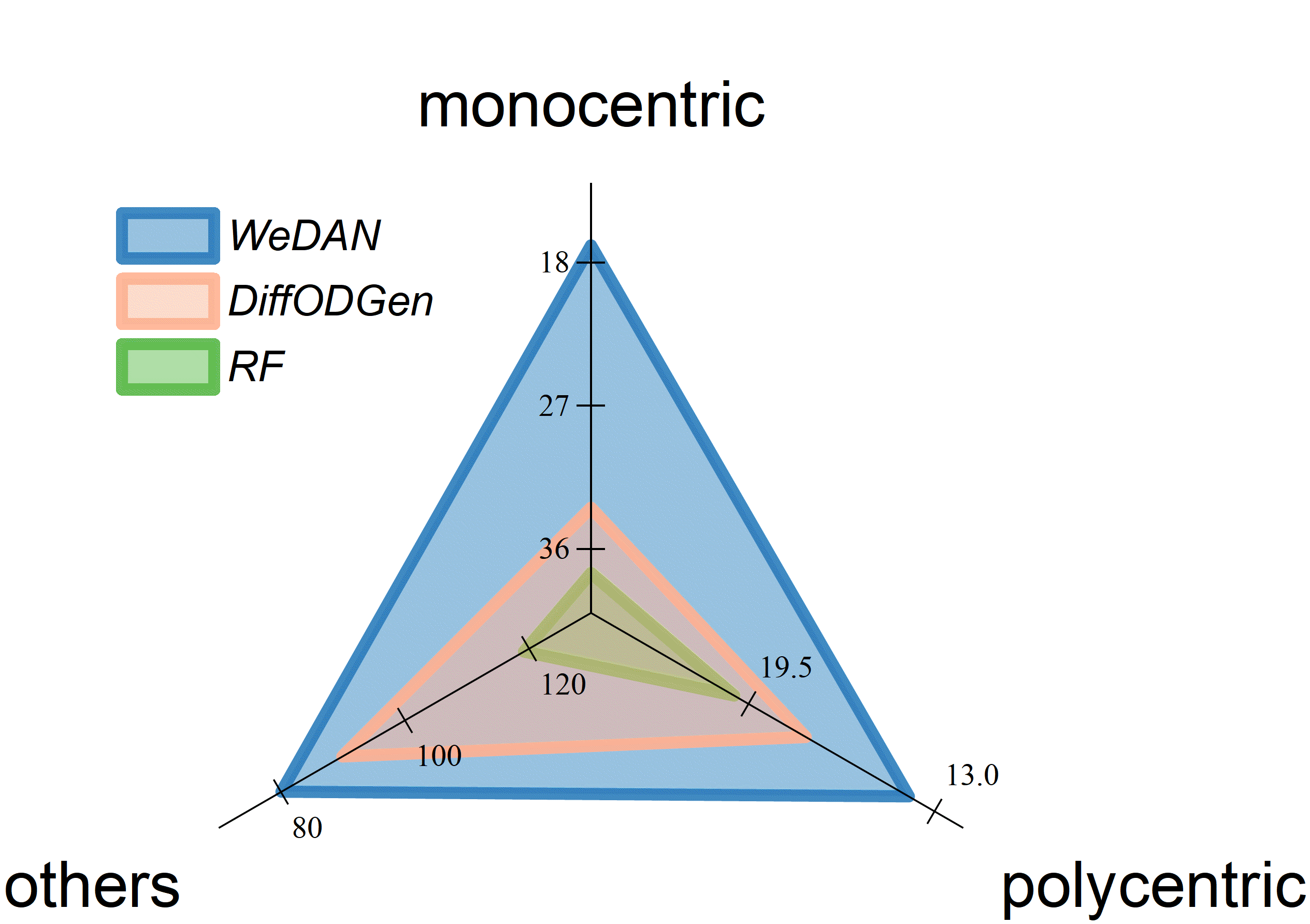}
    }
    \vspace{-0.2cm}
    \caption{Performance comparison across areas of different sizes and structures.}
    \vspace{-0.4cm}
    \label{fig:radars}
\end{figure}
We divided the test areas into six groups based on the number of regions and into three groups based on structure, and the results are shown in Figure~\ref{fig:radars}. We find that when trained under the new paradigm, WeDAN can consistently achieve optimal performance across areas of all sizes and structures. Polycentric areas often have a larger size and more complex pattern, as they develop satellite towns based on the original monocentric structure. Therefore, polycentric areas are more challenging to deal with. However, our model still achieved optimal performance in CPC. Larger areas tend to have more structured layouts, so smaller areas mostly fall into the 'others' category, resulting in better metrics for this category. While DiffODGen is specifically designed for large areas, our method can also enhance its performance by 11.1\% on CPC and 33.3\% on RMSE thanks to the various massive training data. Generative models demonstrate better adaptability to different structures of areas. And our method averagely improves the performance by 34.9\% on RMSE on the monocentric and polycentric areas.

\vspace{-0.2cm}
\subsection{Analysis the Commonalities Captured Across Various Areas}

\begin{figure}[t]
    \centering
    \subfigure[From monocentric to polycentric areas.]{
        \label{fig:mono2poly}
        \includegraphics[width=0.22\textwidth]{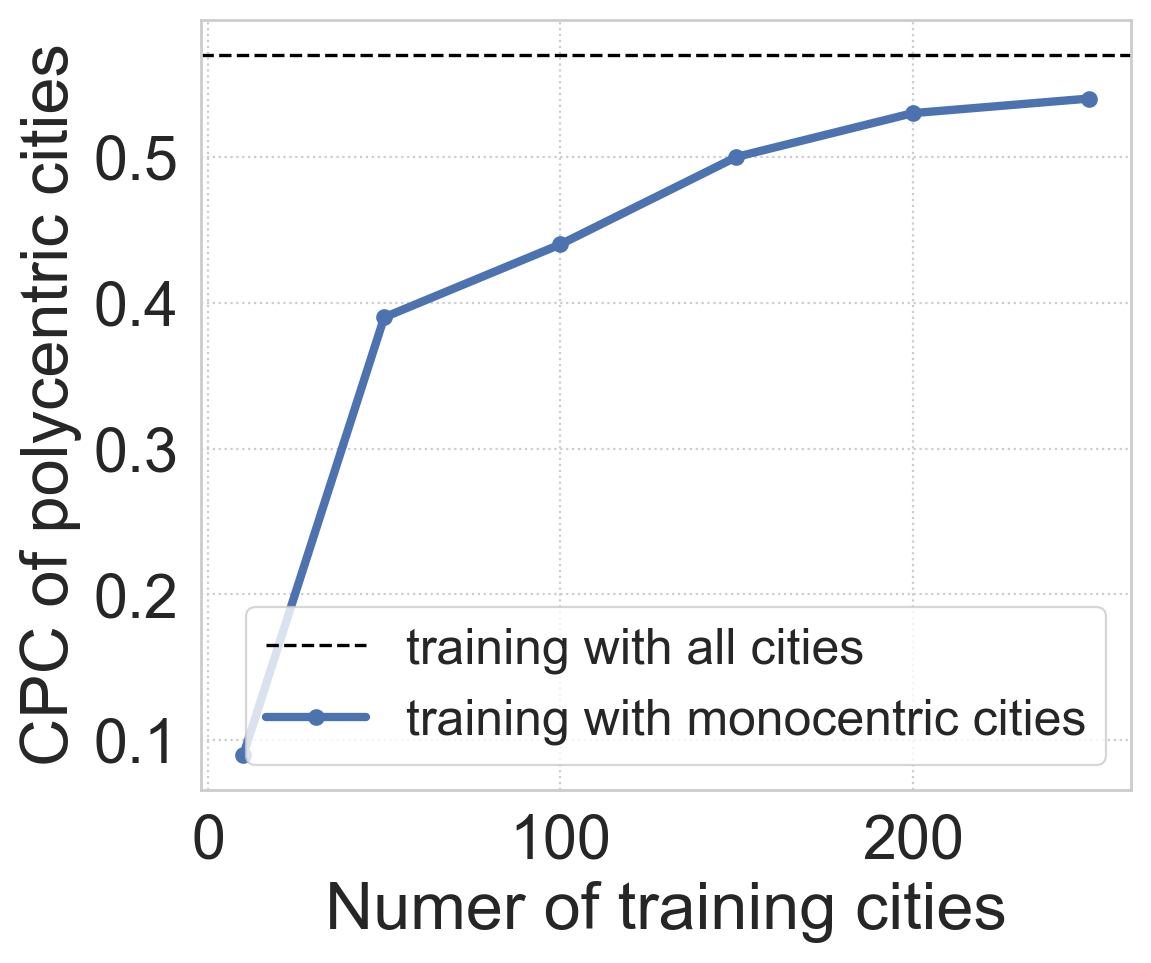}
    }
    \subfigure[From polycentric to monocentric areas.]{
        \label{fig:poly2mono}
        \includegraphics[width=0.22\textwidth]{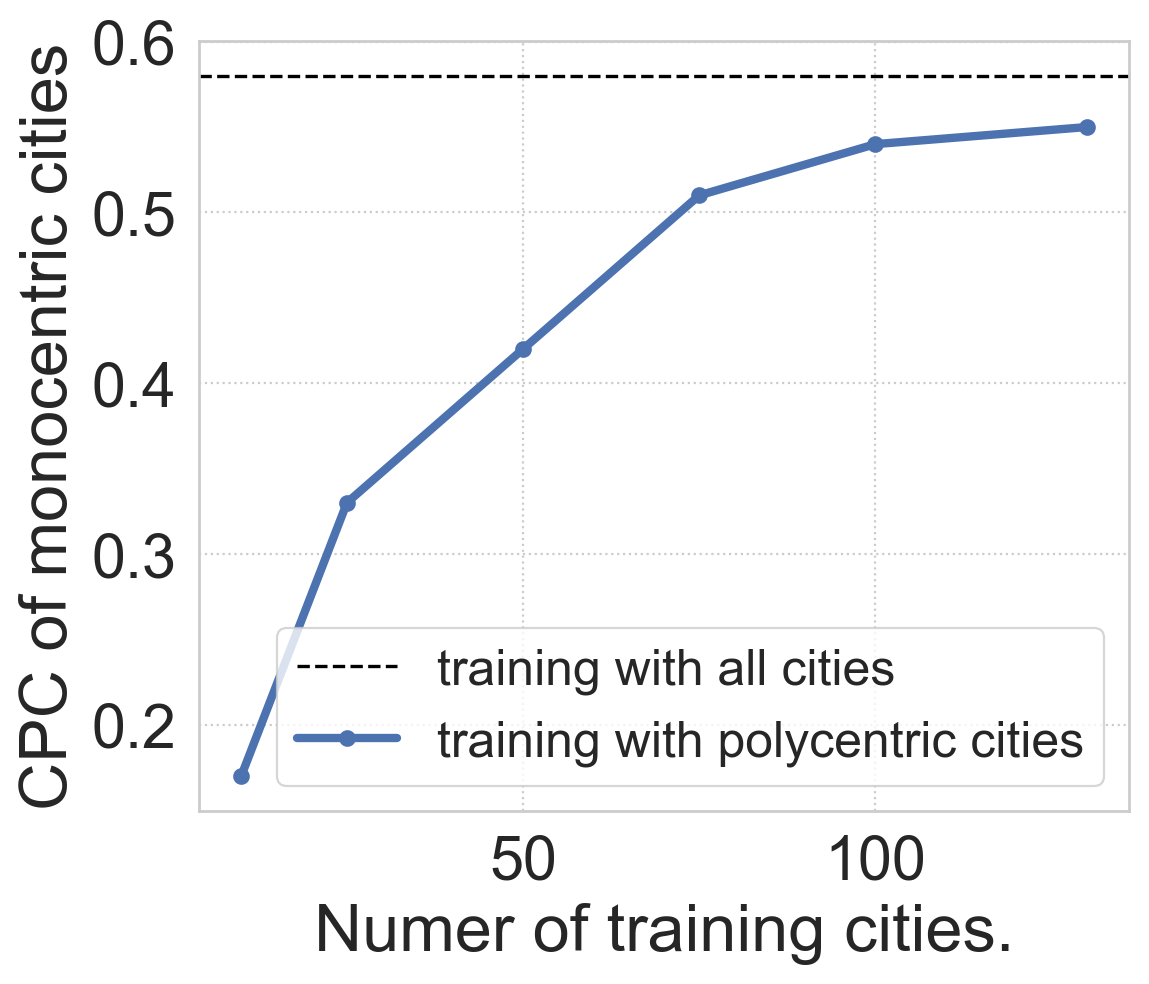}
    }
    \subfigure[From small to large areas.]{
        \label{fig:small2big}
        \includegraphics[width=0.22\textwidth]{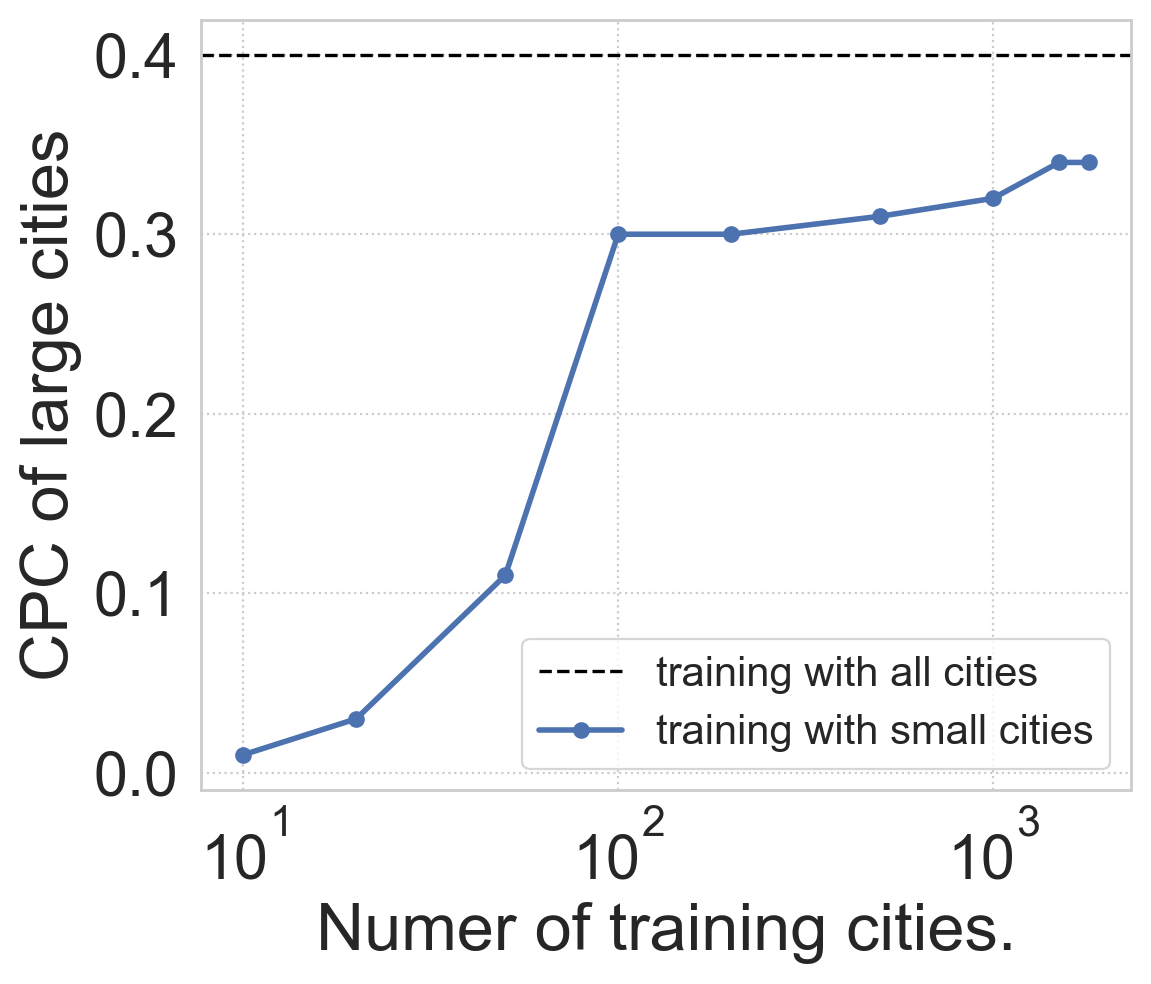}
    }
    \subfigure[From large to small areas.]{
        \label{fig:big2small}
        \includegraphics[width=0.22\textwidth]{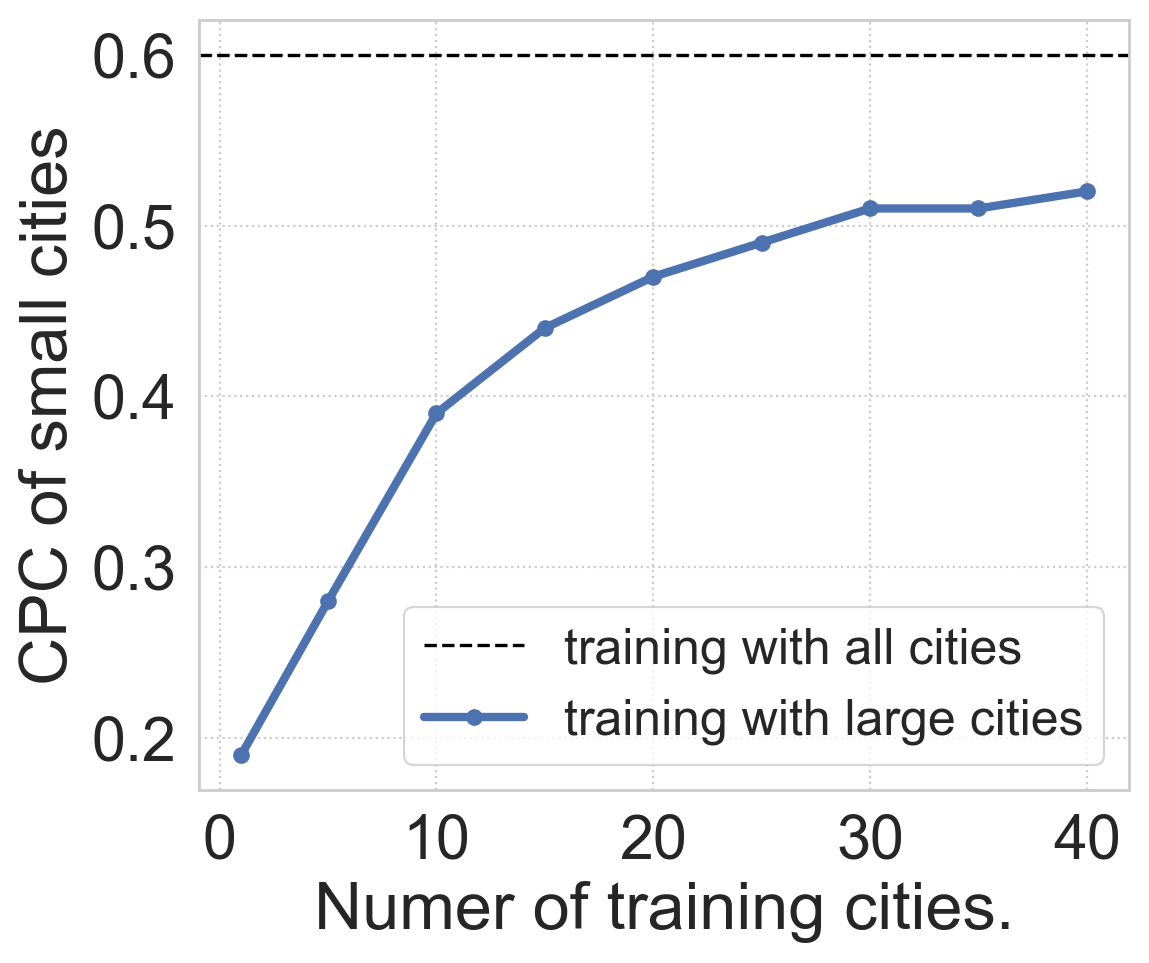}
    }
    \vspace{-0.3cm}
    \caption{Analysis the dependencies captured across areas with different sizes and structures. The small areas consist of less than 100 regions, and the large areas consist of more than 500 regions. The black dash line represents the performance of training with all types of areas.}
    \vspace{-0.4cm}
    \label{fig:cross_dependencies}
\end{figure}

We conducted in-depth analysis of the dependencies captured across areas of different sizes and structures~\cite{xu2023urban} in the new paradigm. Specifically, we utilize areas with varying sizes and structures to mutually serve as training and testing sets, thereby validating the capture of commonalities across areas. The results, as illustrated in Figure~\ref{fig:cross_dependencies}, we find that there are commonalities across areas with different sizes and structures, allowing for a certain degree of mutual transfer between them. Experiments have shown that a performance of 89.7\% can be achieved solely through cross-type transfer learning and applications. Large areas contains more information about flows. Therefore, achieving a performance of 86.7\% can be accomplished with only a small number of training large areas. Training the model with a diverse range of areas can enhance its generalizability, allowing it to achieve good performance across various areas. This indicates the validity of the novel paradigm.

\section{Conclusion} \label{sec:Conclusion}
In this work, we introduce a large-scale benchmark dataset for commuting OD matrix generation, which includes 3,233 areas around the United States. With this dataset, we aim to facilitate the development of more generalizable commuting OD matrix generation models that can capture various patterns of distinct areas all around. Supported by our proposed dataset, we propose a novel paradigm, which considers the whole area combined with its commuting OD matrix as an attributed directed weighted graph and generates the weighted edges based on the node attributes. Based on the benchmark dataset, we compare the proposed paradigm with existing models, including physical models, element-wise predictive models, and matrix-wise generative models. The results show that the proposed paradigm can achieve the optimal. This can give rise to a new research direction from the graph perspective in this field.

\clearpage


\bibliographystyle{plain}
\bibliography{reference}

\clearpage

\appendix

\section{Additional Information About the New Paradigm} \label{apdx:paradigm}

\subsection{Denoising Network}
During each step in the reverse denoising process, the denoising network predicts the small Gaussian noise $\epsilon$ to be removed, based on the current noisy state.  
We adopt the transformer-based neural network structure as the backbone, which has been proven to have strong learning and generalization capabilities across various domains. 

\begin{figure}[h]
    \centering
    \includegraphics[width=0.49\textwidth]{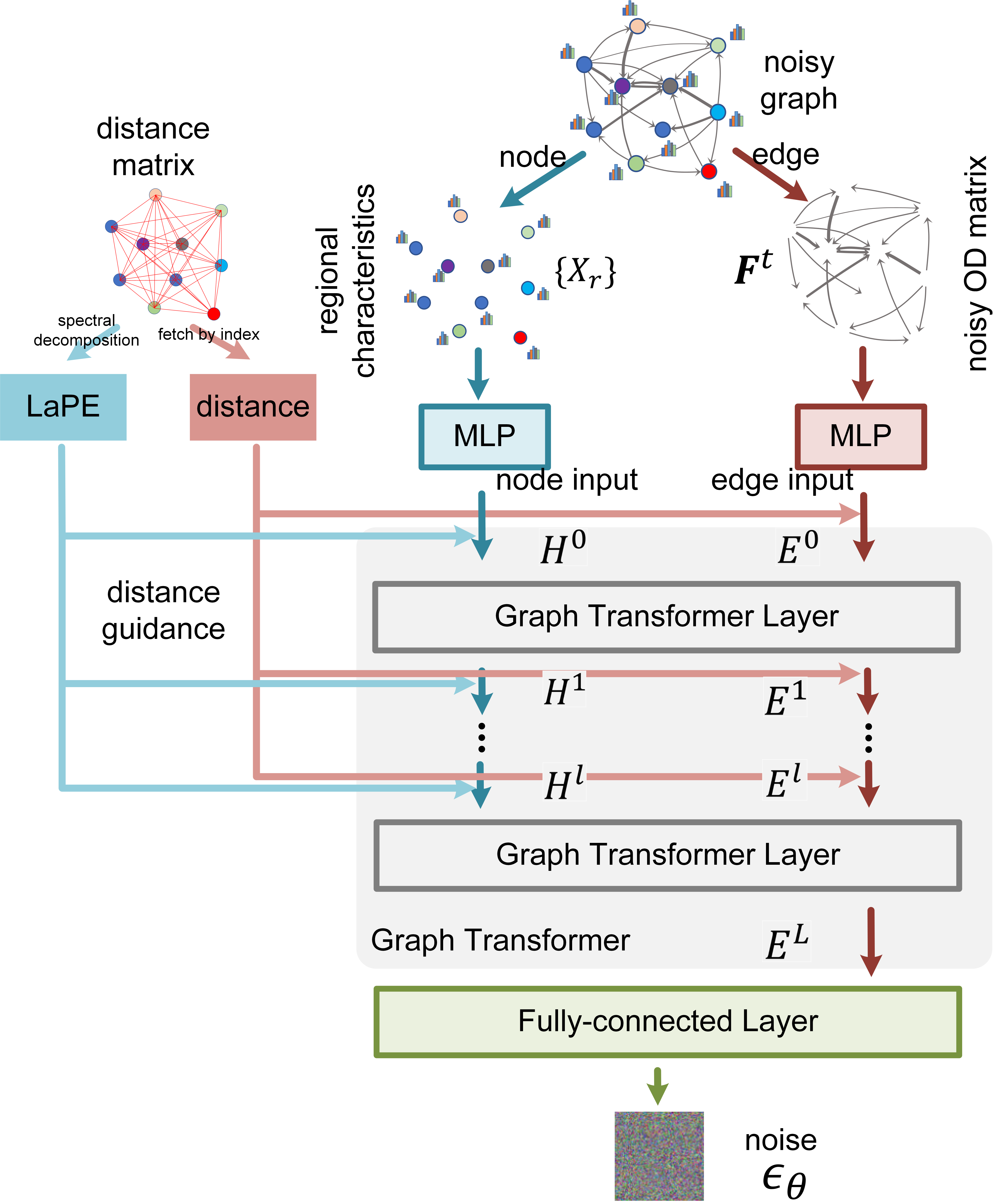}
    \caption{The architecture of denoising neural network $\theta$ of our graph diffusion model.}
    \label{fig:neuralnet}
\end{figure}

As illustrated in Figure~\ref{fig:neuralnet}, the backbone of our denoising network is the graph transformer~\cite{dwivedi2020generalization}. It accepts inputs at both the node and edge levels, captures graph features, and then outputs noise predictions at the edge level. The characteristics of each region serve as node inputs, and the noisy OD matrix at the current state provides the edge inputs. They are processed separately through their respective Multilayer Perceptrons~(MLPs) and then fed into the graph transformer. The graph transformer consists of a series of layers. In each layer, every node computes attention weights with all other nodes through the self-attention mechanism and aggregates information from all other nodes based on these weights. 
To model the dependencies between nodes and edges, the weights computed through self-attention are fused with edge features using Feature-wise Linear Modulation~(FiLM)~\cite{perez2018film}, resulting in the final attention weights. Simultaneously, the calculated attention information is also used to combine with the original edge features, serving as the new edge features for subsequent computations in the next layer of denoising network. Moreover, after the aggregation of node and edge information, the data passes through a feed-forward network. The computations within each graph transformer layer can be described by the following formula.

\begin{equation}
    \begin{split}
    \begin{aligned}
        h_i^{l+1} &= O_h^l \Vert_{k=1}^K ( \sum_{r_j \in \mathcal{N}_{r_i}} \alpha_{ij}^{k,l} V^{k,l} h_j^l ),\\
        e_{ij}^{l+1} &= O_e^l \Vert_{k=1}^K ( a_{ij}^{k,l} ), \\
        \alpha_{ij}^{k,l} &= softmax_j (a_{ij}^{k,l}), \\
        a_{ij}^{k,l} &= ( \frac{ Q^{k,l} h_i^l \cdot K^{k,l} h_j^l }{ \sqrt{d_k} } ) + W^{k,l} e_{ij}^l ,
    \end{aligned}
    \end{split}
\end{equation}
where $h_i^l$ and $e_{ij}^l$ are the node and edge features at the $l$-th layer, respectively. $Q^{k,l}$, $K^{k,l}$, and $V^{k,l}$ are the query, key, and value matrices of the $k$-th attention head at the $l$-th layer. $W^{k,l}$ is the weight matrix of the $k$-th attention head at the $l$-th layer. $O_h^l$ and $O_e^l$ are the output MLPs of the node and edge features at the $l$-th layer. $d_k$ is the dimension of the query and key vectors. $K$ is the number of attention heads. $\mathcal{N}_{v_i}$ is the set of neighbor nodes that are connected to node $v_i$.

After the layers, the final edge features are fed into the fully-connected layer to predict the noise.

\textbf{Distance-based guidance.} To fully utilize the association between spatial interactions $ \{ d_{ij} | r_i \ \text{and} \ r_j  \in \mathcal{R} \} $ and the OD matrix $\mathbf{F}$, we have designed node and edge levels distance-based conditional guidance to direct the denoising generation. As shown in Figure~\ref{fig:neuralnet}, we perform spectral decomposition on the distance matrix to obtain $N$ Laplacian eigenvectors, which named distance-based Laplacian position encodings~(d-LaPEs) are used to encode the specific position of each region in the planar urban space. Subsequently, the node features and edge features, before being inputted into each graph transformer layer, are combined with the corresponding d-LaPEs and distances.

\textbf{Log-Transform.} Existing theoretical works have discovered scaling behaviors in human mobility~\cite{jiang2016timegeo,saberi2017complex,saberi2018complex}, namely that many properties follow the power law distribution. To enable our model to better capture the heterogeneity of OD flow distributions across different cities, we use log-transform to preprocess and post-process OD flows. The calculations are as follows.
\vspace{-0.1cm}
\begin{equation}
    \begin{split}
    \begin{aligned}
        \dot{F_{ij}} & = \log (F_{ij} + 1), \\
        F_{ij} & = \exp (\dot{F_{ij}}) - 1.
    \end{aligned}
    \end{split}
\end{equation}
where $\dot{F_{ij}}$ is the log-transformed OD flow, which is used to train the denoising network. The generated $\dot{F_{ij}}$, after inverse transformation, yields the real size of OD flows $F_{ij}$.

\vspace{-0.2cm}
\subsection{Training and Generation}

We use the simple loss from DDPM~\cite{ho2020denoising} to train the denoising networks in our attributed graph diffusion model. This involves minimizing the Mean Squared Error~(MSE) between the noise predicted by the denoising network and the noise from the forward diffusion process. The calculation of this loss is as follows.

\vspace{-0.5cm}
\begin{equation}
    \begin{split}
    \begin{aligned}
        \mathcal{L} = \mathbb{E}_{t,\epsilon \sim \mathcal{N}(0,\mathbf{I})} \left[ \| \mathbf{\epsilon} - \mathbf{\epsilon}_\theta(\mathbf{F}^t,t,\mathcal{C_\mathcal{R}}) \|_2^2 \right]
    \end{aligned}
    \end{split}
\end{equation}
where $\| \dot \|$ denotes the $L-2$ norm. The training algorithm is shown in Algorithm~\ref{alg:train}. The training and sampling methods are detailed in Appendix~\ref{apdx:train}.

\section{Additional Experimental Details} \label{apdx:expdet}

\subsection{Training Algorithm of Graph Denoising Diffusion} \label{apdx:train}

The trained denoising network can be utilized in conjunction with the reverse denoising process to generate the OD matrix for new cities, which lack any OD flow information, using their spatial characteristics. We adopt the sampling algorithm from Denoising Diffusion Implicit Models~(DDIM)~\cite{song2020denoising} to facilitate more efficient data generation. The sampling algorithm is shown in Algorithm~\ref{alg:generation}.

\renewcommand{\algorithmicrequire}{ \textbf{Input:}}
\renewcommand{\algorithmicensure}{ \textbf{Output:}}

\begin{algorithm}[h]
    \caption{Training of the Graph Diffusion Model}
    \label{alg:train}
    \begin{algorithmic}[1]
    \REQUIRE ~~\\
    Graphs $\mathcal{G}_{train}$ that constructed from the data collected from the cities in training set
    
    \ENSURE ~~\\
    Learned noise prediction neural networks $\theta$.
    
    \STATE Sample a graph $G$ from $\mathcal{G}_{train}$
    \STATE Sample $t \sim \mathcal{U} (1,2,...,T)$ 
    \STATE Sample $\mathbf{\epsilon} \sim \mathcal{N}(0, \mathbf{I}) $
    \STATE $loss \Longleftarrow \left \| \left[ \mathbf{\epsilon} - \mathbf{\epsilon}_{\theta} (\sqrt{\bar{\alpha}^t} F + \sqrt{1-\bar{\alpha}^t} \mathbf{\epsilon}, t, \mathcal{C_\mathcal{R}}) \right] \right \|^2 $
    \STATE optimizer.step($loss$)
    \end{algorithmic}
\end{algorithm}

\begin{algorithm}[h]
    \caption{OD Matrix Generation through Trained Graph Diffusion Model}
    \label{alg:generation}
    \begin{algorithmic}[1]
    \REQUIRE ~~\\
    Spatial characteristics $\mathcal{C_\mathcal{R}}$ of a new city \\
    Trained denoising network $\theta$ \\
    Length $\tau$ of sub-sequence in DDIM sampling
    \ENSURE ~~\\
    OD matrix $\mathbf{F}$ of that new city.
    
    \STATE Sample $\mathbf{F}^T$ $\sim$ $\mathcal{N}(0, \mathbf{I})$
    \STATE $\Delta t = \frac{T}{\tau}$ 
    \FOR{$t = T, T-\Delta t, ..., 1$}
        \STATE $\mathbf{F}^{t-\Delta t} \leftarrow \frac{1}{\sqrt{\alpha_t}} \left( \mathbf{F}^t - \frac{1-\alpha_t}{\sqrt{1-\bar{\alpha}_t}}\epsilon_\theta(\mathbf{F}^t, t, \mathcal{C_\mathcal{R}}) \right)$
    \ENDFOR
    \RETURN $\mathbf{F}^0$
    \end{algorithmic}
\end{algorithm}

\subsection{Architecture of Denoising Graph Transformer} \label{apdx:architecture}

\subsection{Baselines} \label{apdx:baselines}

Within the first category are two conventional methodologies:
\begin{itemize}[itemsep=2pt,topsep=2pt,parsep=0pt,leftmargin=*]
    \item \textbf{Gravity Model (GM)}~\cite{barbosa2018human} is inspired by the gravitation in physics, positing that the OD flow is directly proportional to the populations of the origin and the destination, and inversely proportional to the distance between them.
    \item \textbf{Random Forest (RF)}~\cite{pourebrahim2019trip} stands out as a tree-based machine learning algorithm known for its robustness, demonstrating commendable results in generating OD flows.
\end{itemize}

The second category encompasses two cutting-edge deep-learning approaches tailored for OD flow modeling:
\begin{itemize}[itemsep=2pt,topsep=2pt,parsep=0pt,leftmargin=*]
    \item \textbf{Deep Gravity Model (DGM)}~\cite{simini2021deep} integrates deep learning into traditional gravity models to calculate flows by estimating the distribution probabilities across various regions. We have adapted this model to generate OD flow volumes directly.
    \item \textbf{Geo-contextual Multitask Embedding Learning (GMEL)}~\cite{liu2020learning} leverages graph neural networks (GNNs) to aggregate neighboring information for each region. This process enhances the spatial characteristic representation of the regions in a city, which contributes to the refinement of regional embeddings and augments precision.
\end{itemize}

Lastly, to verify the effectiveness of our design in generating realistic mobility networks, namely the OD matrices, we compare our model with two sophisticated deep generative models:
\begin{itemize}[itemsep=2pt,topsep=2pt,parsep=0pt,leftmargin=*]
    \item \textbf{NetGAN}~\cite{bojchevski2018netgan} is a GAN-style framework that recreates realistic network architectures by generating random walks that mirror the distribution of walks extracted from real networks. We have tailored it to construct directed and weighted graphs, i.e., OD matrices.
    \item \textbf{DiffODGen}~\cite{rong2023complexity} employs a cascaded diffusion model specifically for large cities, leveraging the sparsity of the mobility network to separately model the topology of the graph and the weights given edges, achieved SOTA results in large cities.
\end{itemize}

\subsection{Evaluation Metrics} \label{apdx:metrics}

The specific calculation methods for each metric are as follows.
\begin{equation}
    RMSE{\;}={\;}\sqrt{{\frac{1}{|\textbf{F}|}}{\sum\nolimits_{r_i,r_j\in{\mathcal{R}}}{||}{{{\textbf{F}}_{ij}}-{\hat{\textbf{F}}_{ij}}{{||}_2^2}}}},
\end{equation}
\begin{equation}
    NRMSE{\;}={\;}{ RMSE / \sqrt{ \frac{1}{N^2} \sum\nolimits_{r_i,r_j\in{\mathcal{R}}} || F_{ij} - \bar{F}_{ij} ||_2^2 } } ,
\end{equation}
\begin{equation}
    CPC={ 2\!\! \sum_{r_i,r_j\in{\mathcal{R}}}    \min(\mathbf{F}_{i,j}, \hat{\mathbf{F}}_{i,j}) / ({\sum_{r_i,r_j\in{\mathcal{R}}}{\!\mathbf{F}_{ij}}+ \sum_{r_i,r_j\in{\mathcal{R}}}{\!\hat{\mathbf{F}}_{ij}} }) },
\end{equation}
\begin{equation}
    JSD{\;}={\;} {{\textbf{KL}(\textbf{P}_{\mathbf{F}}||\textbf{P}_{\hat{\mathbf{F}}}) +  \textbf{KL}(\textbf{P}_{\hat{\mathbf{F}}}||\textbf{P}_{\mathbf{F}})}/{2}},
\end{equation}
where the $\bar{\mathbf{F}}$ denotes the mean of elements in OD matrix $\mathbf{F}$, $\textbf{KL}$ means Kullback–Leibler divergence, and $\mathbf{P}$ denotes the empirical probability distribution. The inflow is determined by totaling all flows entering each region, while the outflow is calculated by summing up all flows leaving each region.

\subsection{Parameter Settings} \label{apdx:parameters}
The graph transformer employs 4 layers with each having 32 hidden dimensions. We utilize 1000 diffusion steps in the diffusion model, following a cosine noise scheduler as suggested by Nichol et al.~\cite{nichol2021improved}. The denoising network are optimized using the AdamW optimizer~\cite{loshchilov2017decoupled}, with a learning rate set at 1e-3. Our method and DiffODGen both sample 10 times during generation and take the average as the final generated result.

For the gravity model, we adopt the approach outlined by Barbosa et al.~\cite{barbosa2018human}, which involves four fitting parameters. In the random forest algorithm, the number of estimators is set to 100. The DGM~\cite{simini2021deep} is stacked by 10 layers with 64 hidden dimensions in each layer, while GNN-based models are designed with 3 layers and 64 channels all. The hyper-parameters for the denoising networks in two cascaded diffusion models of DiffODGen are aligned with our methodology.

All the selection of hyper-parameters is based on the validation set and trade off between the performance and computational resources.

\subsection{Details for Reproducibility} \label{apdx:reproducibility}
The computational resources we used to conduct the experiments are as follows: a server with a Intel(R) Xeon(R) Platinum 8358P CPU @ 2.60GHz with 128 cores. The server is equipped with 1TB of RAM and 8 NVIDIA A100 GPUs. For all the experiments in the paper, we run 5 trials and report the average results.

\end{document}